\documentstyle[11pt,aaspp4]{article}

\slugcomment{Submitted to the PASP.}

\begin{document}
\title{{\sl Hubble Space Telescope\/} WFPC2 Imaging\footnote{Based on observations
made with the NASA/ESA {\sl Hubble Space Telescope}, obtained in part from
the data archive of the Space Telescope Science Institute, which is
operated by the Association of Universities for Research in Astronomy,
Inc., under NASA contract NAS 5-26555.} of SN 1979C and Its Environment}

\author{Schuyler D.~Van Dyk\altaffilmark{2}, Chien Y.~Peng\altaffilmark{3}, Aaron J.~Barth\altaffilmark{4}, and Alexei V.~Filippenko} 
\authoremail{vandyk@ipac.caltech.edu}

\authoremail{cyp@as.arizona.edu}

\authoremail{abarth@astro.berkeley.edu}


\affil{Department of Astronomy, University of California, Berkeley, CA  94720-3411}
\authoremail{alex@wormhole.berkeley.edu}

\author{Roger A.~Chevalier}
\affil{Department of Astronomy, University of Virginia, Charlottesville, VA 22903}
\authoremail{rac5x@virginia.edu}

\author{Robert A.~Fesen}
\affil{Department of Physics \& Astronomy, Dartmouth College, Hanover, NH 03755}
\authoremail{fesen@snr.dartmouth.edu}

\author{Claes Fransson}
\affil{Stockholm Observatory, SE-133 36 Saltsj\"obaden, Sweden}
\authoremail{claes@astro.su.se}

\author{Robert P.~Kirshner}
\affil{Harvard-Smithsonian Center for Astrophysics, 60 Garden St., Cambridge, MA 02138}
\authoremail{kirshner@cfa.harvard.edu}

\and 

\author{Bruno Leibundgut}
\affil{European Southern Observatory, Karl Schwarzschild Strasse 2, Garching D-85748, 
Germany}
\authoremail{bleibundgut@eso.org}

\altaffiltext{2}{Current address: IPAC/Caltech, Mailcode 100-22,
Pasadena, CA 91125.}

\altaffiltext{3}{Current address: Steward Observatory, University of Arizona,
Tucson, AZ 85721.}

\altaffiltext{4}{Current address: Harvard/Smithsonian Center for Astrophysics,
60 Garden St., Cambridge, MA 02138.}

\begin{abstract}
The locations of supernovae in the local stellar and gaseous environment in galaxies 
contain important clues to their progenitor stars.  As part of a program to study
the environments of supernovae using {\sl Hubble Space Telescope\/} ({\sl HST}) 
imaging data, we have examined the environment of the Type II-L SN 1979C in 
NGC 4321 (M100).  We place more rigorous constraints on the mass of the 
SN progenitor, which may have had a mass $M \approx 17$--18 $M_{\sun}$.  
Moreover, we have recovered and measured the brightness of SN 1979C,
$m=23.37$ in F439W ($\sim B$; $m_B({\rm max}) = 11.6$), 17 years after explosion.
\end{abstract}


\section{Introduction}

A primary goal of supernova research is an understanding of the progenitor stars and 
explosion mechanisms of the different types of supernovae (SNe).  In the absence of direct
information about the progenitor stars, scrutiny of the host galaxies and local environments
of SNe continues to yield valuable clues to their nature.  Previous ground-based studies of
SN host galaxies and environments have primarily concentrated on statistical results.  
Investigation of the local stellar and gaseous environments of SNe can, in favorable cases, 
yield useful constraints on the ages and masses of progenitor stars.  However, most studies
of this kind have been hampered by the limited spatial resolution of ground-based 
observations.  The superior angular resolution of the {\sl Hubble Space Telescope\/} 
({\sl HST}) offers the potential for greater understanding of SN environments.

Presumably caused by the core collapse of massive 
stars, Type II SNe (SNe~II) have been associated with a young stellar population (e.g., 
\cite{van92a}).  SNe~II all exhibit hydrogen in their optical spectra, but the strength and
profile of the H$\alpha$ line vary widely among these objects (e.g., \cite{sch96}).
At late times, SNe~II are dominated by the strong H$\alpha$ emission line.
Photometrically, SNe II are subclassified into ``plateau'' (SNe II-P) and ``linear'' 
(SNe II-L), based on the shape of their light curves (\cite{bar79}; \cite{dog85}).
As part of a larger survey of SNe environments using {\sl HST\/} WFPC2 archive data
(\cite{van98}),
we have examined the environment of the SN II-L 1979C in NGC 4321 (M100).

SN 1979C was discovered on 1979 April 19 by Johnson (1979) at around magnitude 12.
Optical spectra first showed a featureless continuum, which later slowly evolved to
exhibit strong H$\alpha$ emission, but with weak or no
P Cygni absorption (Panagia et al.~1980; Branch et al.~1981; Barbon et al.~1982;
Schlegel 1996).  Photometrically, SN 1979C was unusually blue near maximum (which
occurred on or about 1979 April 15, at $m_B({\rm max}) = 11.6$; \cite{dev81})
and declined in brightness in a way characteristic of SNe II-L (e.g., \cite{bar82}).
The SN was also extraordinarily luminous, at $M_B({\rm max}) \approx -20$ (e.g., \cite{you89}),
making it the brightest SN II yet observed.

SN 1979C is also a bright late-time radio source (\cite{wei86}, 1991).  The radio sphere of
the SN was resolved by VLBI observations (\cite{bar85}).  Weiler et al.~(1991)
showed that its radio emission is consistent with a red supergiant progenitor star which
initially had a mass $\gtrsim$13 $M_{\odot}$.  Periodic undulations in the evolution of 
the radio emission led Weiler et al.~(1992) to model the progenitor as possibly being in
a detached eccentric binary system with a less massive companion, similar to the VV
Cephei systems. Hydrodynamical simulations by Schwarz \& Pringle (1996) confirm that this
binary system model is feasible for SN 1979C.  The radio emission has been declining in a 
relatively normal fashion, until, after almost two decades, it has ceased declining and
may be rising again at all radio frequencies (Montes et al.~1998).

According to the Chevalier (1982) scenario, the radio emission is best modelled as the 
interaction of the SN shock wave with relatively high-density circumstellar matter, set
up by a constant mass-loss rate, constant velocity wind from the red supergiant progenitor.
Chevalier \& Fransson (1994) have shown from hydrodynamical considerations for SN 1979C 
that X-ray radiation from the SN shock front can be absorbed by a shell formed via 
radiative cooling at the reverse shock.  This
gives rise to a low-ionization optical spectrum in the ejecta and in the shocked shell, 
while high-ionization lines are formed in the freely expanding SN ejecta.  This implies
late-time optical emission at H$\alpha$, [O~I], [O~III], and other lines.  Fesen \& 
Matonick (1993) from the ground, and Fesen et al.~(1998) with {\sl HST\/}, have indeed
observed strong late-time line emission in the optical and UV, and interpret this as
the result of the SN shock-circumstellar gas interaction.

The site of SN 1979C was imaged on two sets of {\sl HST\/} data we have analyzed, with details of 
the separate analyses provided below.  Based on its radio position (\cite{wei91}), 
Van Dyk et al.~(1996) found this SN to be associated with a faint H~II region of radius
$1{\farcs}5$, situated below the bright southern spiral arm in M100 (see also \cite{fes93}).
From the {\sl HST\/} images, the SN is seen to have occurred in or near a small cluster of 
stars, some of which presumably contribute to the ionization of that H~II region.  We have also
recovered and measured the brightness of the SN in both the broad-band and narrow-band {\sl HST\/}
images obtained in 1996 July, 17 years after explosion.

\section{Analysis}

For the first time, a number of resolved stars are seen in the SN 1979C environment on both
sets of {\sl HST\/} images discussed below.  We employed point spread function (PSF) fitting
photometry of these stars performed by DAOPHOT (\cite{ste87}) and ALLSTAR
within IRAF\footnote{IRAF (Image Reduction and Analysis 
Facility) is distributed by the National Optical Astronomy Observatories, which are operated
by the Association of Universities for Research in Astronomy, Inc., under cooperative 
agreement with the National Science Foundation.}.  Stars were located on the images using 
DAOFIND, with a detection threshold of 3$\sigma$, determined by the
gain and read noise parameters for the image.  Because of the lack of isolated stars
of sufficient signal-to-noise ratio on the images we analyzed, it was impossible to 
build a good model PSF from field stars on these images.  Instead, we employed the 
{\sl Tiny Tim\/} routine (Krist 1995) to produce an artificial PSF.  The PSF-fitting 
photometry resulted in color-magnitude diagrams (CMDs) for the stars in the SN environment.
Throughout this paper we express the magnitudes and colors in the WFPC2 flight magnitude 
system.  To analyze these diagrams, we used the theoretical isochrones for solar metallicity 
from Bertelli et al.~(1994), in order to constrain the ages and masses of the
stars.  (We have converted the Johnson-Cousins $UBVRI$ magnitudes and colors for the
isochrones into WFPC2 flight system magnitudes and colors, using the transformations given in
Table 2 of Van Dyk et al.~1998.)
We have assumed the distance modulus to M100 ($m-M=31.04$, or $d=16.1$ Mpc)
measured using Cepheids by Ferrarese et al.~(1996).

\subsection{{\sl HST\/} Extragalactic Distance Scale Key Project Images}

The data obtained from the {\sl HST\/} archive, consisting of 12 F555W and four F814W 
cosmic-ray-split pairs of images, were originally part of the Extragalactic Distance
Scale Key Project (\cite{fer96}).  
Details of these observations are given in Table 1.

To take advantage of the very high signal-to-noise ratio made possible by the long
exposures, we created images combined in each band from all the
individual image sets.  Before combining, we first registered the individual images 
using the cross-correlation method XREGISTER in IRAF, as well as using positions 
of the brightest stars in each image.  Rotation of the individual images was negligible.  
The accuracy of the image alignment was estimated to be well within 0.3 pixels.  The combined
F555W image is shown in Figure 1.  One can see the small star cluster within the 1\arcsec\
radius error circle for the SN's position.  (The radio position is accurate to $\pm 0{\farcs}2$, 
but we have
assigned a more conservative error, due to the larger uncertainty in the world coordinates
of the {\sl HST\/} images.)

The resolved stars in the associated cluster have separations of only $\sim$3 to 5 pixels 
peak-to-peak.
Because of crowding, and loss of resolution in the shifted and undersampled images, it
was necessary to identify and separate stars by eye in the associated star cluster.  All
identified stars were removed by PSF subtraction; stars which remained after subtraction
were then added to the star list.  We subsequently used this list derived from the combined
image in each band and applied it within ALLSTAR to each of the individual images that went
into forming the combination.  
The crowding and variable background required that the sky be determined with a small 
local annulus of $0{\farcs}3$ inner radius and $0{\farcs}4$ width.  Trial and error
dictated the best fitting radius, from $0{\farcs}14$ to $0{\farcs}18$.

The instrumental magnitudes determined from individual images were weight-averaged
to yield final magnitudes.  
To stay consistent with the methods and results in Ferrarese et al.~(1996), instrumental
magnitudes were converted to apparent magnitudes using the zeropoints from Hill et 
al.~(1998).

In Figure 2 we show the CMD for the SN 1979C environment (the star cluster nearest the 
positional error circle), which results from the photometry on the individual images, along with
photometry of the summed images for stars not found on the individual images.  The brightest star
on the diagram is most likely SN 1979C itself (see next subsection).
We cannot accurately estimate the reddening for this environment 
from this diagram, but it may be appreciable.  Branch et al.~(1981) estimated that
the reddening to SN 1979C was $E(B-V) \simeq 0.15$ mag, but this could have been largely 
circumstellar.  On the other hand, it is also possible that the underlying
cluster may be reddened by more than this.  
We show in Figure 2 the reddened
isochrones from Bertelli et al.~(1994) with solar metallicity, 
reddening of $E(B-V)=0.15$ mag, and a distance modulus of ($m-M=31.04$),
after transforming the isochrones to the WFPC2 flight system.

\subsection{{\sl HST\/} Project GO 6584 Images}

In addition to the Key Project images, the environment of SN 1979C was observed 
as part of GO Project 6584, whose aim was to study the interaction of SNe with their
circumstellar environments.  Table 1 lists the details of these observations, which were not
nearly as deep as the combined Key Project images, but covered a wider range in wavelength,
including the F658N (H$\alpha$) narrow band, and also were centered on the PC1 chip, affording 
higher spatial resolution than the Key Project images.  In Figure 3 we show the F555W and 
F658N images.  The arrow in each figure points to what is very likely SN 1979C, which was
still optically bright in 1996 July, especially at H$\alpha$, consistent with the results
of Fesen \& Matonick (1993) for 1991 and 1992.  The numbering in Figure 3a corresponds to
the stars listed in Table 2, for which we provide a comparison with our photometry of the
Key Project images discussed above.  (The errors for the magnitudes given in Table 2 are
the formal errors provided by ALLSTAR.)
One can see that the photometry for the two datasets
agrees reasonably well; slight discrepancies can be explained in terms of the higher spatial
resolution of the GO 6584 images {\it versus\/} the higher signal-to-noise ratio of the Key 
Project images.

We can measure the H$\alpha$ line flux from our F658N image in Figure 3b.  Using 
Equation 11 in Holtzman et al.~(1995) to convert a point-source count rate into flux,
we find that the total H$\alpha$ flux on 1996 July 29 UT for SN 1979C through the F658N filter
was $4.7 \times 10^{-16}$ erg cm$^{-2}$ s$^{-1}$.  Fesen \& Matonick (1993; see also
Fesen et al.~1998) detect a broad-line
and narrow-line spectral components at H$\alpha$.  It is thought that the broad-line 
flux is produced by the interaction of the SN shock with the presupernova circumstellar matter
(\cite{che94}), 
which is also the source for the radio emission from SN 1979C (\cite{wei91}).  The 
unresolved narrow-line component is from the associated H~II region, seen by Van Dyk et
al.~(1996).  In the WFPC2-PC1 image the H~II emission from this region is seen mostly to the 
east of the SN (Figure 3b), and therefore likely contributes very little to the F658N
flux.  However, the broad-line component is broader than the F658N bandpass by a factor
$\sim$5.  Taking this into account, the flux we derive here agrees with the 
$2.5 \times 10^{-15}$ erg cm$^{-2}$ s$^{-1}$ measured by Fesen \& Matonick (1993).
Recently, Montes et al.~(1998) have found that the radio emission has ceased declining and 
is possibly rising again.  This could lead to an increase in
the optical emission-line flux as well.

The photometry was conducted, as above, using an appropriate TinyTim PSF in ALLSTAR.
For SN 1979C, we find the following filter magnitudes:  
$m_{\rm F336W}=23.24 \pm  0.09$, $m_{\rm F439W}=23.37 \pm 0.04$, $m_{\rm F555W}=22.15 \pm 0.02$,
$m_{\rm F675W}=20.88 \pm 0.03$, and $m_{\rm F814W}=21.05 \pm 0.04$; also,
$m_{\rm F658N}=21.11 \pm 0.04$, using the synthetic zeropoint in Holtzman et al.~(1995).
The agreement between $m_{\rm F675W}$ and $m_{\rm F658N}$ indicates that most
of the emission in this broad band is from the H$\alpha$ line (see also \cite{fes93}), 
and that the continuum contribution to $m_{\rm F658N}$ is small.

In Figure 4 we show the CMDs for the associated cluster.  We correct the isochrones,
assuming the same distance modulus and reddening as we did in the previous subsection.
The bright red star with relatively small uncertainties is most likely the SN.  

\section{Results and Conclusions}

We conclude from both datasets that a noncoeval mixture of
young populations exists in the environment of SN 1979C, with very young ($\sim$4--6 Myr) blue 
stars to older ($\sim$20 Myr) red supergiants.  Most of the stars immediately surrounding the 
SN appear to be massive main-sequence turnoff stars and a red supergiant star with ages $\sim$10
Myr.  If the progenitor of SN 1979C was a member of this population, then its age was also
$\sim$10 Myr.  (If $E[B-V]$ to the SN were greater by, e.g., 0.5 mag, this age could be in error, 
i.e., too old, by $\sim$3--4 Myr.)
Based on the SN's radio emission (\cite{wei86}, 1991), as well as its optical
light curve, it is likely that 
the progenitor was a red supergiant star.  Assuming solar metallicity for the models in 
Bertelli et al.~(1994), this implies a mass for the progenitor of $\sim$17--18 $M_{\sun}$.

This mass estimate
is consistent with the constraint on the mass ($\gtrsim$13 $M_{\sun}$) made by Weiler et 
al.~(1991), based on an estimate of the amount of presupernova mass loss.  Constraints
on the masses of SN progenitors have been placed on SNe 1987A, 1980K, and 1968L, based on 
the properties of their stellar environments.  For SN 1987A, Romaniello et al.~(1998) and Van Dyk,
Hamuy, \& Mateo (1998) find that the most
luminous blue stars in the SN's environment have ages 10--12 Myr, coeval to the SN progenitor
(if the progenitor had mass $19\pm3$ $M_{\sun}$; \cite{arn89}).
For the SN II-L 1980K Thompson (1982) set an upper limit of 18 $M_{\sun}$ on the progenitor 
mass from a plate taken before explosion.  However, this result could be affected
by the progenitor being enshrouded in dust.  For the SN II-P 1968L, Barth et al.~(1996) find that 
its progenitor may have had an initial mass of $\gtrsim$25--30 $M_{\sun}$, if it were coeval with
the star clusters in its environment.

This is the first time, however,
other than for SN 1987A (\cite{rom98}), that the mass of a SN progenitor has been 
estimated from the stellar populations in its environment measured from {\sl HST\/} images.

\acknowledgements

Financial suport for this work was provided by NASA through Grant Nos.~AR-5793, AR-6371, and 
GO-6043 from the Space Telescope Science Institute, which is operated by AURA, Inc., under
NASA Contract No.~NAS 5-26555; we also acknowledge
NSF grant AST-9417213. 
We appreciate fruitful consultations with Craig Sosin and Adrienne Cool, and discussions with 
Rob Kennicutt regarding the Key Project data for M100.  
This research has made use of the NASA/IPAC Extragalactic Database (NED) which is 
operated by the Jet Propulsion Laboratory, California Institute of Technology, under 
contract with the National Aeronautics and Space Administration.

\clearpage

\begin{deluxetable}{cccc}
\def\phmm{\phm{$-$}}
\tablenum{1}
\tablecolumns{7}
\tablewidth{7in}
\tablecaption{Summary of Data for the SN 1979C Environment}
\tablehead{\colhead{Camera/Chip} & \colhead{Filter} & \colhead{UT Date} & \colhead{Exposure (s)}}
\startdata
WFPC2/WF-2\tablenotemark{a} & F555W & 1994 Apr 23--Jun 19 & 21600 \nl
         & F814W &                     & 7200 \nl
WFPC2/PC-1\tablenotemark{b} & F336W & 1996 Jul 29 & 2600 \nl
                          & F439W &             & 2400 \nl
                          & F555W &             & 800 \nl
                          & F675W &             & 1000 \nl
                          & F814W &             & 1200 \nl
                          & F658N &             & 3900 \nl
\enddata
\tablenotetext{a}{These data were obtained from the {\sl HST\/} Archive.}
\tablenotetext{b}{These images were taken as part of {\sl HST\/} program GO 6584.}
\end{deluxetable}

\clearpage

\begin{deluxetable}{ccccc}
\def\phmm{\phm{$-$}}
\tablenum{2}
\tablecolumns{5}
\tablewidth{6in}
\tablecaption{Comparison of Photometry for the SN 1979C Field}
\tablehead{\colhead{} & \multicolumn{2}{c}{Key Project Images} & \multicolumn{2}{c}{GO 6043 Images} \nl
\colhead{Star} & \colhead{$m_{\rm F555W}$} & \colhead{$m_{\rm F814W}$} 
& \colhead{$m_{\rm F555W}$} & \colhead{$m_{\rm F814W}$}} 
\startdata
1\tablenotemark{a} & $22.20{\pm}.01$ & $21.17{\pm}.01$ & $22.15{\pm}.02$ & $21.05{\pm}.04$ \nl 
2 & $25.11{\pm}.04$ & $23.51{\pm}.05$ & $24.78{\pm}.19$ & $23.56{\pm}.10$ \nl 
3 & $23.94{\pm}.02$ & $23.51{\pm}.05$ & $24.26{\pm}.10$ & $23.89{\pm}.11$ \nl 
4 & $24.23{\pm}.01$ & $23.57{\pm}.03$ & $24.47{\pm}.10$ & $23.63{\pm}.05$ \nl 
5 & $24.32{\pm}.02$ & $24.28{\pm}.07$ & $24.72{\pm}.13$ & $24.27{\pm}.13$ \nl 
6 & $24.42{\pm}.03$ & $24.23{\pm}.07$ & $24.74{\pm}.10$ & $24.39{\pm}.17$ \nl 
\enddata
\tablenotetext{a}{This star is most likely SN 1979C.}
\end{deluxetable}
\clearpage

\begin{figure}
\figurenum{1}
\caption{The environment of SN 1979C in NGC 4321 (M100), as seen on a F555W WFPC2 image,
summed from the individual images obtained by the {\sl HST\/} Key Project.  
The position of the SN is at the center of the circle, which represents the 1\arcsec\
radius uncertainty in position.  The bright star within the circle is most likely SN 1979C 
itself (see Figure 3).  Orientation of the image is north up, east to the left.}
\end{figure}

\begin{figure}
\figurenum{3}
\caption{The environment of SN 1979C in NGC 4321 (M100), as seen on WFPC2 images
obtained as part of GO 6584, in the {\it (a)} F555W band and {\it (b)} F658N band.  
The arrow in each figure points to what is likely the SN itself.  The numbering on
{\it (a)} indicates the stars listed in Table 2.  Orientation of the image is north up, 
east to the left.}
\end{figure}

\clearpage

\begin{figure}
\figurenum{2}
\plotone{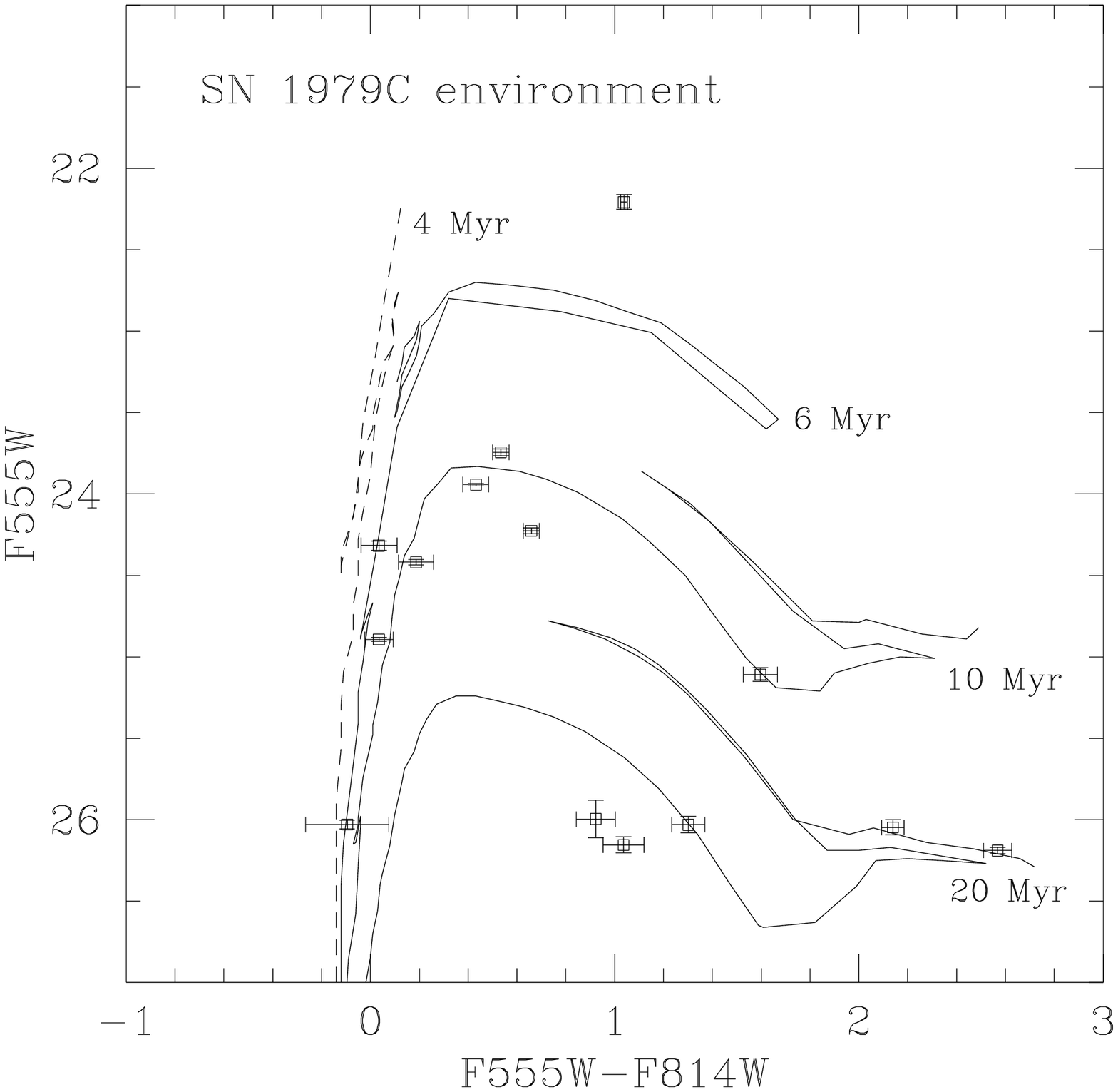}
\caption{The color-magnitude diagram for the environment of SN 1979C in NGC 4321 (M100),
based on {\sl HST\/} Key Project archival data.  Also shown are isochrones from Bertelli
et al.~(1994), reddened by $E(B-V)=0.15$ mag (Branch et al.~1981), corrected for the distance
modulus to M100 ($m-M=31.04$) from Ferrarese et al.~(1996), and transformed to the WFPC2 flight 
system colors.}
\end{figure}

\clearpage

\begin{figure}
\figurenum{4}
\plotone{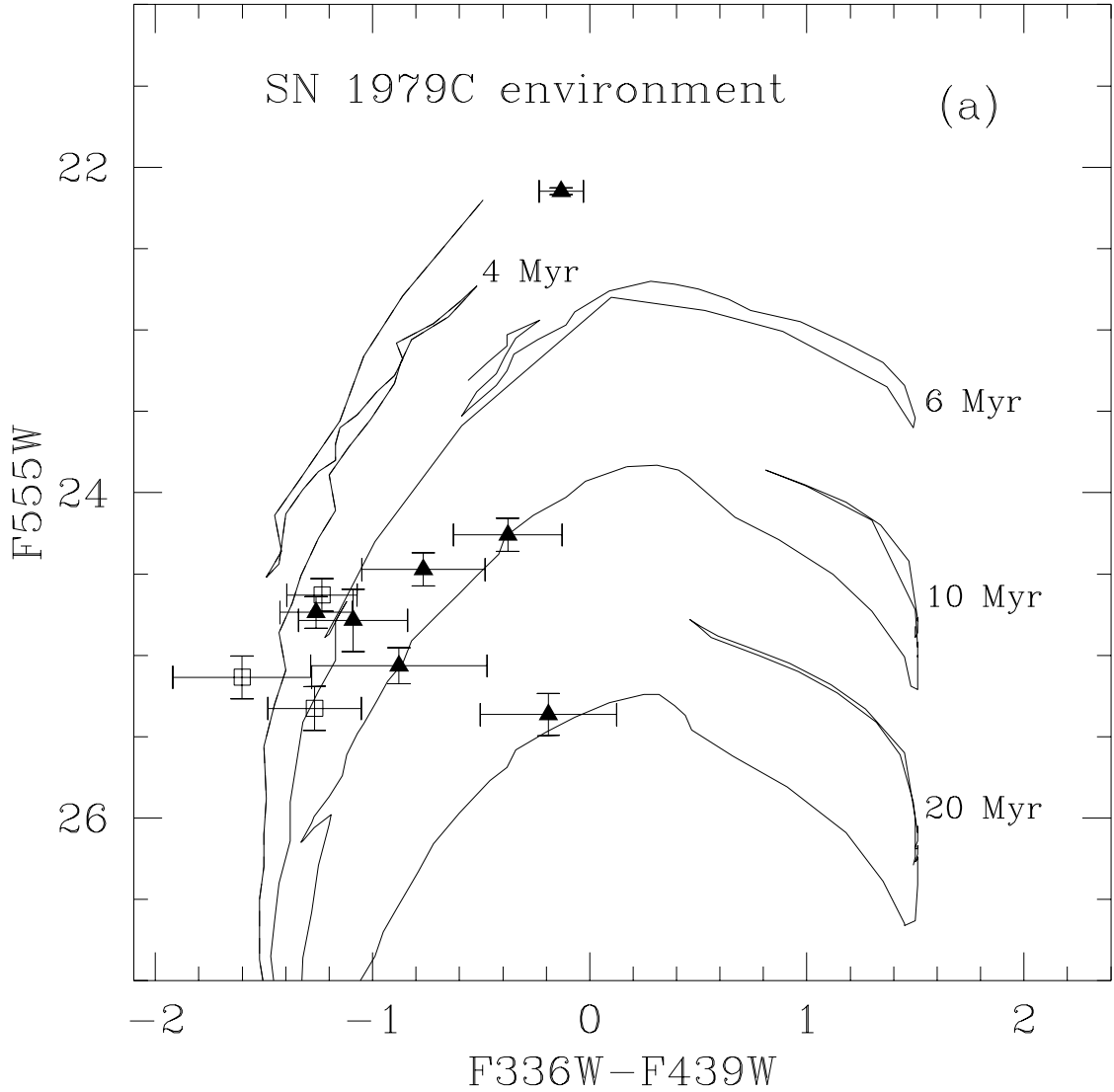}
\caption{The color-magnitude diagrams for the environment of SN 1979C in NGC 4321 (M100),
obtained as part of {\sl HST\/} program GO 6584.  The {\it filled triangles\/} are data
for which magnitudes were obtained in all five broad bands; the {\it open squares\/}
are data for which magnitudes were obtained in at least two broad bands.
Also shown are isochrones from Bertelli
et al.~(1994), reddened by $E(B-V)=0.15$ mag (Branch et al.~1981), corrected for the distance
modulus to M100 ($m-M=31.04$; Ferrarese et al.~1996), and  
transformed to the WFPC2 flight system colors.}
\end{figure}

\clearpage

\begin{figure}
\figurenum{4}
\plotone{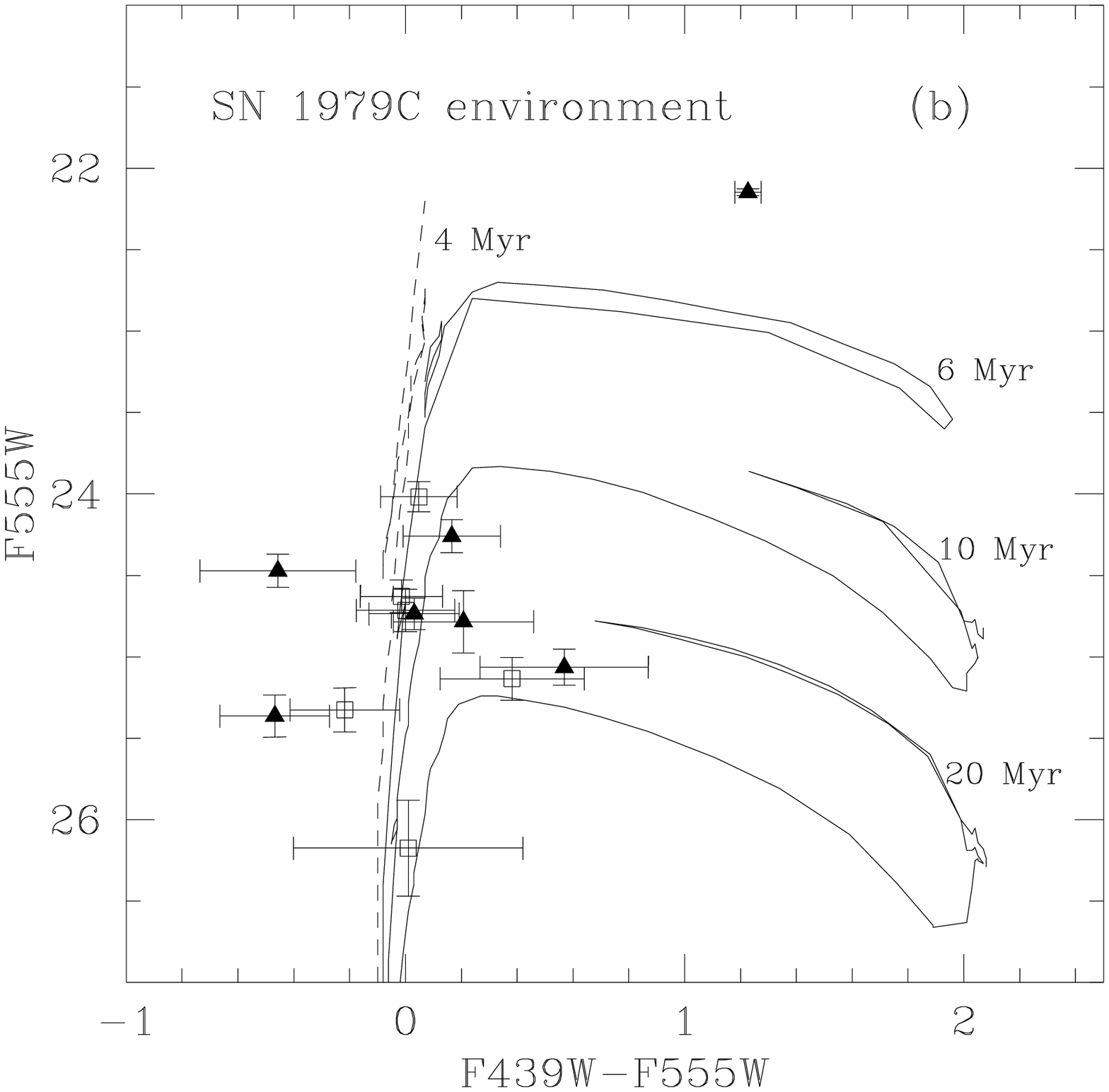}
\caption{(Continued.)}
\end{figure}

\clearpage

\begin{figure}
\figurenum{4}
\plotone{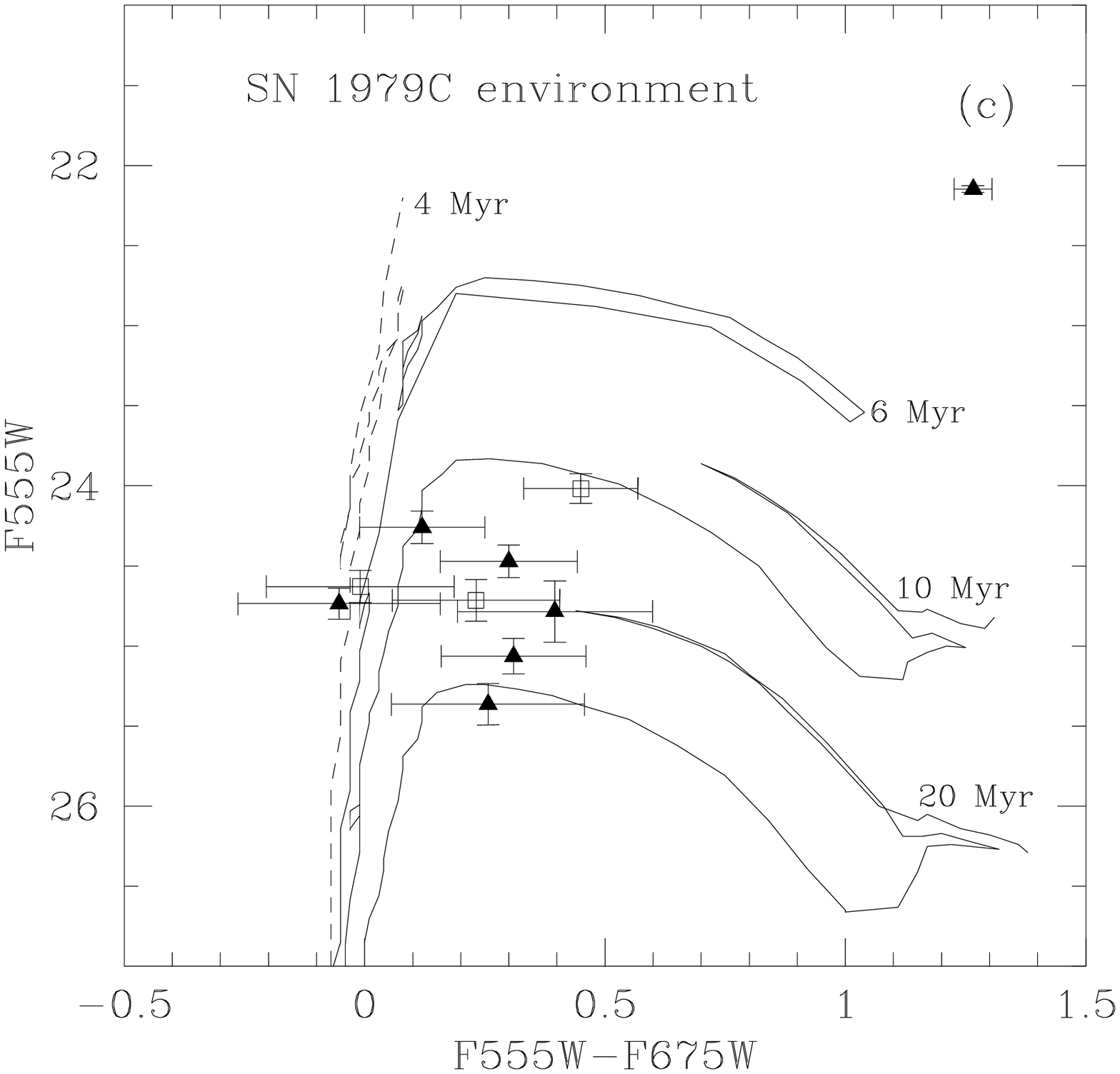}
\caption{(Continued.)}
\end{figure}

\clearpage

\begin{figure}
\figurenum{4}
\plotone{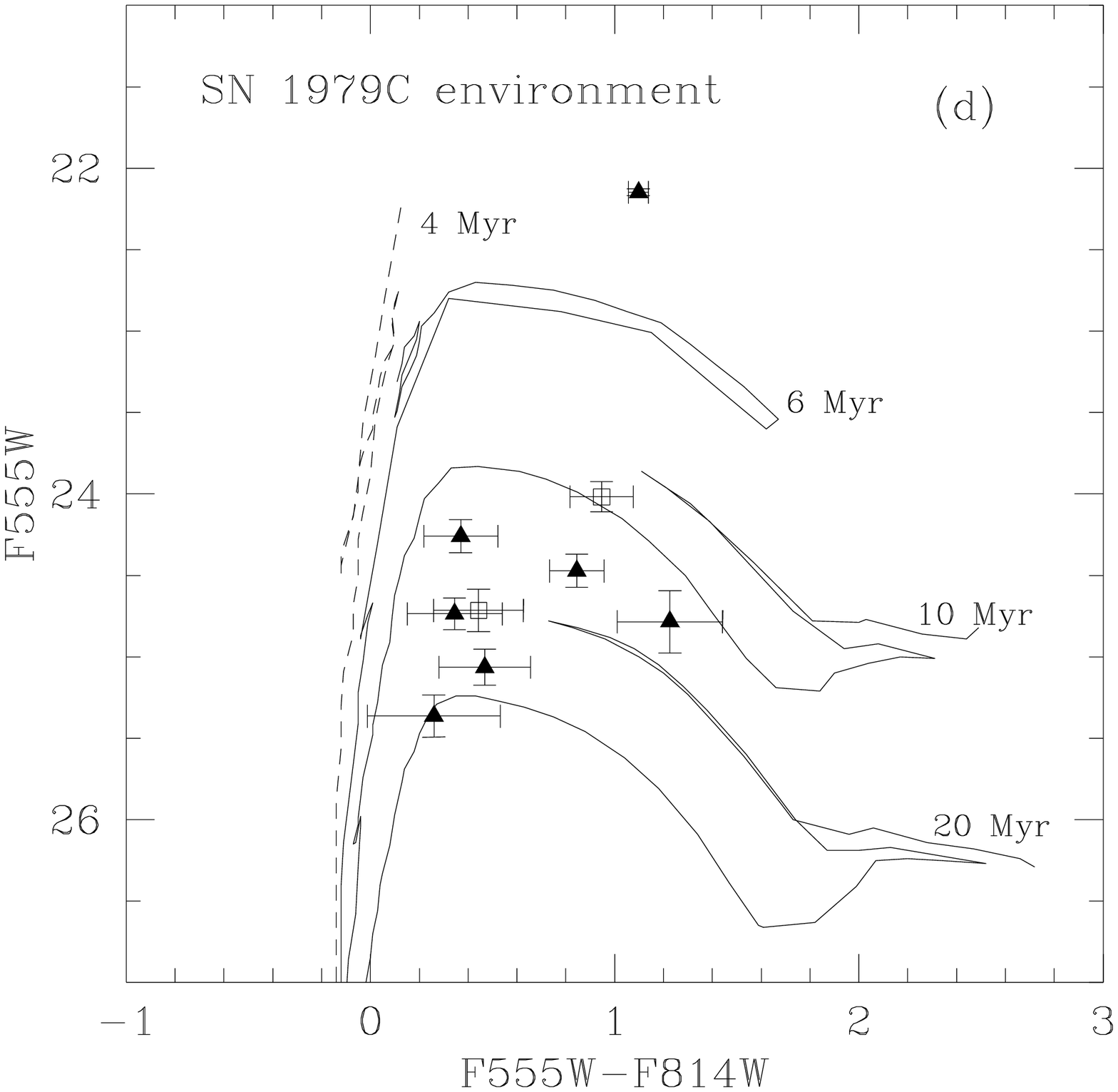}
\caption{(Continued.)}
\end{figure}


\begin{thebibliography}{}
\bibitem[Arnett et al.~1989]{arn89} Arnett, D.~W., Bahcall, J.~N., Kirshner, R.~P.,
\& Woosley, S.~E. 1989, \araa, 27, 629
\bibitem[Barbon, Ciatti, \& Rosino 1979]{bar79} Barbon, R., Ciatti, F., \& Rosino, L.
1979, \aap, 72, 287
\bibitem[Barbon et al.~1982]{bar82} Barbon, R., Ciatti, F., Rosino, L., Ortolani,
S., \& Rafanelli, P. 1982, \aap, 116, 43
\bibitem[Bartel et al.~1985]{bar85} Bartel, N., Shapiro, I.~I., Gorenstein, M.~V.,
Gwinn, C.~R., \& Rogers, A.~E.~E. 1985, Nature, 318, 25
\bibitem[Bertelli et al.~1994]{ber94} Bertelli, G., Bressan, A., Chiosi, C.,
Fagotto, F., \& Nasi, E. 1994, \aaps, 106, 275
\bibitem[Branch et al.~1981]{bra81} Branch, D., Falk, S.~W., McCall, M.~L., 
Rybski, P., Uomoto, A.~K., \& Wills, B.~J. 1981, \apj, 244, 780
\bibitem[Chevalier 1982]{che82} Chevalier, R.~A. 1982, \apj, 259, 302
\bibitem[Chevalier \& Fransson 1994]{che94} Chevalier, R. A., \& Fransson, C. 1994,
\apj, 420, 268
\bibitem[de Vaucouleurs et al.~1981]{dev81} de Vaucouleurs, G., de Vaucouleurs, A.,
Buta, R., Ables, H.~D., \& Hewitt, A.~V. 1981, \pasp, 93, 36
\bibitem[Doggett \& Branch 1985]{dog85} Doggett, J.~B., \& Branch, D. 1985, \aj, 90,
2303
\bibitem[Ferrarese et al.~1996]{fer96} Ferrarese, L., et al.~1996, \apj, 464, 568
[Erratum: 475, 853 (1997)]
\bibitem[Fesen \& Matonick 1993]{fes93} Fesen, R.~A., \& Matonick, D.~M. 1993, \apj,
407, 110
\bibitem[Fesen et al.~1998]{fes98} Fesen, R.~A., et al.~1998, AJ, submitted
\bibitem[Hill et al.~1998]{hil98} Hill, R. J., et al.~1998, \apj, 496, 648
\bibitem[Johnson 1979]{joh79} Johnson, G.~E. 1979, \iaucirc 3348
\bibitem[Krist 1995]{kri95} Krist, J. 1995, Tiny Tim User's Manual, version 3.0a
\bibitem[Montes et al.~1998]{mon98} Montes, M.~J., Weiler, K.~W., Van Dyk, S.~D.,
Sramek, R.~A., Panagia, N., \& Park, R. 1998, in preparation
\bibitem[Panagia et al.~1980]{pan80} Panagia, N., et al.~1980, \mnras, 192, 861
\bibitem[Romaniello et al.~1998]{rom98} Romaniello, M., Panagia, N., Scuderi, S.,
et al.~1998, in The Magellanic Clouds and Other Dwarf Galaxies, 
ed.~T.~Richtler \& J.~M.~Braun (Shaker Verlag), in press
\bibitem[Schlegel 1996]{sch96} Schlegel, E.~M. 1996, \aj, 111, 1660
\bibitem[Schwarz \& Pringle 1996]{sch96b} Schwarz, D.~H., \& Pringle, J.~E. 1996,
\mnras, 282, 1018
\bibitem[Stetson 1987]{ste87} Stetson, P.~B. 1987, \pasp, 99, 191
\bibitem[Thompson 1982]{tho82} Thompson, L.~A. 1982, \apjl, 257, L63
\bibitem[Van Dyk 1992]{van92a} Van Dyk, S.~D. 1992, \aj, 103, 1788
\bibitem[Van Dyk et al.~1996]{van96} Van Dyk, S.~D., Hamuy, M., \& Filippenko, A.~V.
1996, \aj, 111, 2017
\bibitem[Van Dyk, Hamuy, \& Mateo 1998]{van98b} Van Dyk, S.~D., Hamuy, M., \& Mateo,
M. 1998, in SN 1987A: Ten Years Later, eds.~M.~M.~Phillips \& N.~B.~Suntzeff, 
ASP Conf.~Ser., in press
\bibitem[Van Dyk et al.~1998]{van98} Van Dyk, S.~D., Peng, C.~Y., Barth, A.~J., \& 
Filippenko, A.~V., \aj, submitted
\bibitem[Weiler et al.~1986]{wei86} Weiler, K.~W., Sramek, R.~A., Panagia, N., van der
Hulst, J.~M., \& Salvati, M. 1986, \apj, 301, 790
\bibitem[Weiler et al.~1991]{wei91} Weiler, K.~W., Van Dyk, S.~D., Panagia, N., 
Sramek, R.~A., \& Discenna, J.~L. 1991, \apj, 380, 161
\bibitem[Weiler et al.~1992]{wei92} Weiler, K.~W., Van Dyk, S.~D., Pringle, J.~E.,
\& Panagia, N. 1992, 399, 672
\bibitem[Young \& Branch 1989]{you89} Young, T.~R., \& Branch, D. 1989, \apjl, 342,
L79
\end{thebibliography}
\end{document}